\catcode`\@=11
\expandafter\ifx\csname @iasmacros\endcsname\relax
       \global\let\@iasmacros=\par
\else  \endinput
\fi
\catcode`\@=12


\def\rmb{\seventeenrm}

\def\itb{\twelvebxi}

\def\singlespace{\baselineskip=\normalbaselineskip}
\def\halfspace{\baselineskip=1.5\normalbaselineskip}
\def\doublespace{\baselineskip=2\normalbaselineskip}


\def\AB{\bigskip\parindent=40pt
        \centerline{\bf ABSTRACT}\medskip\halfspace\narrower}
\def\AE{\bigskip\nonarrower\doublespace}
\def\nonarrower{\advance\leftskip by-\parindent
       \advance\rightskip by-\parindent}


\def\boxit#1{\vbox{\hrule\hbox{\vrule\kern3pt
       \vbox{\kern3pt#1\kern3pt}\kern3pt\vrule}\hrule}}

\def\hence{\leavevmode\hbox{\bf .\raise5.5pt\hbox{.}.} }

\def\dalemb#1#2{{\vbox{\hrule height.#2pt
       \hbox{\vrule width.#2pt height#1pt \kern#1pt \vrule width.#2pt}
       \hrule height.#2pt}}}
\def\gtorder{\mathrel{\raise.3ex\hbox{$>$}\mkern-14mu
             \lower0.6ex\hbox{$\sim$}}}
\def\ltorder{\mathrel{\raise.3ex\hbox{$<$}\mkern-14mu
             \lower0.6ex\hbox{$\sim$}}}

\newdimen\fullhsize
\newbox\leftcolumn
\def\twoup{\hoffset=-.5in \voffset=-.25in
  \hsize=4.75in \fullhsize=10in \vsize=6.9in
  \def\fullline{\hbox to\fullhsize}
  \let\lr=L
  \output={\if L\lr
        \global\setbox\leftcolumn=\columnbox\global\let\lr=R \advancepageno
      \else \doubleformat \global\let\lr=L\fi
    \ifnum\outputpenalty>-20000 \else\dosupereject\fi}
  \def\doubleformat{\shipout\vbox{
    \fullline{\box\leftcolumn\hfil\columnbox}\advancepageno}}
  \def\columnbox{\leftline{\vbox{\makeheadline\pagebody\makefootline}}}
  \tolerance=1000 }

\catcode`\@=11                                   



\font\fiverm=cmr5                         
\font\fivemi=cmmi5                        
\font\fivesy=cmsy5                        
\font\fivebf=cmbx5                        

\skewchar\fivemi='177
\skewchar\fivesy='60


\font\sixrm=cmr6                          
\font\sixi=cmmi6                          
\font\sixsy=cmsy6                         
\font\sixbf=cmbx6                         

\skewchar\sixi='177
\skewchar\sixsy='60


\font\sevenrm=cmr7                        
\font\seveni=cmmi7                        
\font\sevensy=cmsy7                       
\font\sevenit=cmti7                       
\font\sevenbf=cmbx7                       

\skewchar\seveni='177
\skewchar\sevensy='60


\font\eightrm=cmr8                        
\font\eighti=cmmi8                        
\font\eightsy=cmsy8                       
\font\eightit=cmti8                       
\font\eightbf=cmbx8                       

\skewchar\eighti='177
\skewchar\eightsy='60


\font\ninei=cmmi9
\font\ninesy=cmsy9

\skewchar\ninei='177
\skewchar\ninesy='60


\font\tenrm=cmr10                         
\font\teni=cmmi10                         
\font\tensy=cmsy10                        
\font\tenex=cmex10                        
\font\tenit=cmti10                        
\font\tensl=cmsl10                        
\font\tenbf=cmbx10                        
\font\tentt=cmtt10                        
\font\tenss=cmss10                        
\font\tensc=cmcsc10                       
\font\tenbi=cmmib10                       

\skewchar\teni='177
\skewchar\tenbi='177
\skewchar\tensy='60

\def\tenpoint{\ifmmode\err@badsizechange\else
       \textfont0=\tenrm \scriptfont0=\sevenrm \scriptscriptfont0=\fiverm
       \textfont1=\teni  \scriptfont1=\seveni  \scriptscriptfont1=\fivemi
       \textfont2=\tensy \scriptfont2=\sevensy \scriptscriptfont2=\fivesy
       \textfont3=\tenex \scriptfont3=\tenex   \scriptscriptfont3=\tenex
       \textfont4=\tenit \scriptfont4=\sevenit \scriptscriptfont4=\sevenit
       \textfont5=\tensl
       \textfont6=\tenbf \scriptfont6=\sevenbf \scriptscriptfont6=\fivebf
       \textfont7=\tentt
       \textfont8=\tenbi \scriptfont8=\seveni  \scriptscriptfont8=\fivemi
       \def\rm{\tenrm\fam=0 }%
       \def\it{\tenit\fam=4 }%
       \def\sl{\tensl\fam=5 }%
       \def\bf{\tenbf\fam=6 }%
       \def\tt{\tentt\fam=7 }%
       \def\ss{\tenss}%
       \def\sc{\tensc}%
       \def\bmit{\fam=8 }%
       \rm\setparameters\setbaselines\fi}


\font\twelverm=cmr12                      
\font\twelvei=cmmi12                      
\font\twelvesy=cmsy10       scaled\magstep1             
\font\twelveex=cmex10       scaled\magstep1             
\font\twelveit=cmti12                            
\font\twelvesl=cmsl12                            
\font\twelvebf=cmbx12                            
\font\twelvett=cmtt12                            
\font\twelvess=cmss12                            
\font\twelvesc=cmcsc10      scaled\magstep1             
\font\twelvebi=cmmib10      scaled\magstep1             
\font\twelvebxi=cmbxti10 scaled\magstep1  

\skewchar\twelvei='177
\skewchar\twelvebi='177
\skewchar\twelvesy='60

\def\twelvepoint{\ifmmode\err@badsizechange\else
       \textfont0=\twelverm \scriptfont0=\eightrm \scriptscriptfont0=\sixrm
       \textfont1=\twelvei  \scriptfont1=\eighti  \scriptscriptfont1=\sixi
       \textfont2=\twelvesy \scriptfont2=\eightsy \scriptscriptfont2=\sixsy
       \textfont3=\twelveex \scriptfont3=\tenex   \scriptscriptfont3=\tenex
       \textfont4=\twelveit \scriptfont4=\eightit \scriptscriptfont4=\sevenit
       \textfont5=\twelvesl
       \textfont6=\twelvebf \scriptfont6=\eightbf \scriptscriptfont6=\sixbf
       \textfont7=\twelvett
       \textfont8=\twelvebi \scriptfont8=\eighti  \scriptscriptfont8=\sixi
       \def\rm{\twelverm\fam=0 }%
       \def\it{\twelveit\fam=4 }%
       \def\sl{\twelvesl\fam=5 }%
       \def\bf{\twelvebf\fam=6 }%
       \def\tt{\twelvett\fam=7 }%
       \def\ss{\twelvess}%
       \def\sc{\twelvesc}%
       \def\bmit{\fam=8 }%
       \rm\setparameters\setbaselines\fi}


\font\fourteenrm=cmr10      scaled\magstep2             
\font\fourteeni=cmmi10      scaled\magstep2             
\font\fourteensy=cmsy10     scaled\magstep2             
\font\fourteenex=cmex10     scaled\magstep2             
\font\fourteenit=cmti10     scaled\magstep2             
\font\fourteensl=cmsl10     scaled\magstep2             
\font\fourteenbf=cmbx10     scaled\magstep2             
\font\fourteentt=cmtt10     scaled\magstep2             
\font\fourteenss=cmss10     scaled\magstep2             
\font\fourteensc=cmcsc10 scaled\magstep2  
\font\fourteenbi=cmmib10 scaled\magstep2  

\skewchar\fourteeni='177
\skewchar\fourteenbi='177
\skewchar\fourteensy='60

\def\fourteenpoint{\ifmmode\err@badsizechange\else
       \textfont0=\fourteenrm \scriptfont0=\tenrm \scriptscriptfont0=\sevenrm
       \textfont1=\fourteeni  \scriptfont1=\teni  \scriptscriptfont1=\seveni
       \textfont2=\fourteensy \scriptfont2=\tensy \scriptscriptfont2=\sevensy
       \textfont3=\fourteenex \scriptfont3=\tenex \scriptscriptfont3=\tenex
       \textfont4=\fourteenit \scriptfont4=\tenit \scriptscriptfont4=\sevenit
       \textfont5=\fourteensl
       \textfont6=\fourteenbf \scriptfont6=\tenbf \scriptscriptfont6=\sevenbf
       \textfont7=\fourteentt
       \textfont8=\fourteenbi \scriptfont8=\tenbi \scriptscriptfont8=\seveni
       \def\rm{\fourteenrm\fam=0 }%
       \def\it{\fourteenit\fam=4 }%
       \def\sl{\fourteensl\fam=5 }%
       \def\bf{\fourteenbf\fam=6 }%
       \def\tt{\fourteentt\fam=7}%
       \def\ss{\fourteenss}%
       \def\sc{\fourteensc}%
       \def\bmit{\fam=8 }%
       \rm\setparameters\setbaselines\fi}


\font\seventeenrm=cmr10 scaled\magstep3          


\newdimen\rp@
\newcount\@basestretchnum
\newskip\@baseskip
\newskip\headskip
\newskip\footskip


\def\setparameters{\rp@=.1em
       \headskip=24\rp@
       \footskip=\headskip
       \delimitershortfall=5\rp@
       \nulldelimiterspace=1.2\rp@
       \scriptspace=0.5\rp@
       \abovedisplayskip=10\rp@ plus3\rp@ minus5\rp@
       \belowdisplayskip=10\rp@ plus3\rp@ minus5\rp@
       \abovedisplayshortskip=5\rp@ plus2\rp@ minus4\rp@
       \belowdisplayshortskip=10\rp@ plus3\rp@ minus5\rp@
       \normallineskip=\rp@
       \lineskip=\normallineskip
       \normallineskiplimit=0pt
       \lineskiplimit=\normallineskiplimit
       \jot=3\rp@
       \setbox0=\hbox{\the\textfont3 B}\p@renwd=\wd0
       \skip\footins=12\rp@ plus3\rp@ minus3\rp@
       \skip\topins=0pt plus0pt minus0pt}


\def\setbaselines{\maxdepth=4\rp@\baselinestretch=\@basestretchnum}


\def\baselinestretch{\afterassignment\@basestretch\@basestretchnum}
\def\@basestretch{%
       \@baseskip=12\rp@ \divide\@baseskip by1000
       \normalbaselineskip=\@basestretchnum\@baseskip
       \baselineskip=\normalbaselineskip
       \bigskipamount=\the\baselineskip
              plus.25\baselineskip minus.25\baselineskip
       \medskipamount=.5\baselineskip
              plus.125\baselineskip minus.125\baselineskip
       \smallskipamount=.25\baselineskip
              plus.0625\baselineskip minus.0625\baselineskip
       \setbox\strutbox=\hbox{\vrule height.708\baselineskip
              depth.292\baselineskip width0pt }}



\def\makeheadline{\vbox to0pt{\baselinestretch=1000
       \vskip-\headskip \vskip1.5pt
       \line{\vbox to\ht\strutbox{}\the\headline}\vss}\nointerlineskip}

\def\makefootline{\baselineskip=\footskip\line{\the\footline}}

\def\big#1{{\hbox{$\left#1\vbox to8.5\rp@ {}\right.\n@space$}}}
\def\Big#1{{\hbox{$\left#1\vbox to11.5\rp@ {}\right.\n@space$}}}
\def\bigg#1{{\hbox{$\left#1\vbox to14.5\rp@ {}\right.\n@space$}}}
\def\Bigg#1{{\hbox{$\left#1\vbox to17.5\rp@ {}\right.\n@space$}}}


\mathchardef\alpha="710B
\mathchardef\beta="710C
\mathchardef\gamma="710D
\mathchardef\delta="710E
\mathchardef\epsilon="710F
\mathchardef\zeta="7110
\mathchardef\eta="7111
\mathchardef\theta="7112
\mathchardef\iota="7113
\mathchardef\kappa="7114
\mathchardef\lambda="7115
\mathchardef\mu="7116
\mathchardef\nu="7117
\mathchardef\xi="7118
\mathchardef\pi="7119
\mathchardef\rho="711A
\mathchardef\sigma="711B
\mathchardef\tau="711C
\mathchardef\upsilon="711D
\mathchardef\phi="711E
\mathchardef\chi="711F
\mathchardef\psi="7120
\mathchardef\omega="7121
\mathchardef\varepsilon="7122
\mathchardef\vartheta="7123
\mathchardef\varpi="7124
\mathchardef\varrho="7125
\mathchardef\varsigma="7126
\mathchardef\varphi="7127
\mathchardef\imath="717B
\mathchardef\jmath="717C
\mathchardef\ell="7160
\mathchardef\wp="717D
\mathchardef\partial="7140
\mathchardef\flat="715B
\mathchardef\natural="715C
\mathchardef\sharp="715D


\def\err@badsizechange{%
       \immediate\write16{--> Size change not allowed in math mode, ignored}}

\baselinestretch=1000
\tenpoint

\catcode`\@=12                                   
\twelvepoint
\doublespace
{\nopagenumbers{
\rightline{~~~September, 2006}
\bigskip\bigskip
\centerline{\rmb Normalization of Collisional Decoherence: Squaring the}
\centerline{\rmb Delta Function, and an Independent Cross-Check}
\medskip
\centerline{\itb Stephen L. Adler
}
\centerline{\bf Institute for Advanced Study}
\centerline{\bf Princeton, NJ 08540}
\medskip
\bigskip\bigskip
\leftline{\it Send correspondence to:}
\medskip
{\singlespace\leftline{Stephen L. Adler}
\leftline{Institute for Advanced Study}
\leftline{Einstein Drive, Princeton, NJ 08540}
\leftline{Phone 609-734-8051; FAX 609-924-8399;
email adler@ias.edu}}
\bigskip\bigskip
}}
\vfill\eject
\pageno=2
\AB
We show that when the Hornberger--Sipe calculation of collisional
decoherence is carried out with the squared delta function a
delta of energy instead of a delta of the absolute value of momentum,
following a method introduced by  Di\'osi, the corrected
formula for the decoherence rate is simply obtained.  The results of
Hornberger and Sipe and of Di\'osi are shown to be in agreement.
As an independent cross-check, we calculate the
mean squared coordinate diffusion of a hard sphere implied by the
corrected decoherence master equation, and show that it agrees precisely
with the same quantity as calculated by a classical Brownian motion analysis.
\AE
\bigskip\bigskip
\vfill\eject
\pageno=3
\centerline{\bf 1.~~Introduction}
\bigskip
The calculation of collisional decoherence was initiated by
Joos and Zeh [1], with generalizations of their
result and corrections to the overall normalization given in papers
of Gallis and Fleming [2], Dodd and Halliwell [3], and Hornberger and
Sipe [4].  An even more general calculation of collisional decoherence
was also given by Di\'osi [5], but this was not known to Hornberger
and Sipe, while noted by Dodd and Halliwell, as well as in the master 
equation papers of Altenm\"uller, M\"uller, and Schenzle [6] and of 
Vacchini [7].  
A difficulty encountered in  refs [1], [2], and [4] is the appearance of
a squared delta function of the absolute value of momentum
in the calculation of the decoherence rate.  To circumvent this,
Hornberger and Sipe carried out a careful wave packet analysis, which gives an
answer smaller by a factor of $2\pi$ than that given by Gallis and
Fleming, and this result is in agreement with experiment [8].

In the course of an alternative derivation of the corrected result, following
the ``traditional approach'' of refs [1] and [2], Hornberger and Sipe
introduce a
rule in which the squared delta function of absolute value of
momentum is evaluated in
terms of an {\it inverse} scattering cross section, which drops out later
in their calculation.  This mixing of kinematic quantities (such as a
squared delta function) and dynamical ones (such as a cross section)
is unconventional, and  Hornberger and Sipe describe this part of their
calculation as speculative.  In Sec. 2 we show that an entirely conventional
completion of the Hornberger--Sipe
calculation is possible, if one follows the method used in the earlier
and more general master equation derivation given by  Di\'osi, and
also used in the derivation of Dodd and Halliwell.  These authors
retain
the delta function of energy that
appears as the ${\cal T}$-matrix coefficient, rather than converting this
delta function to a delta function of the absolute value of momentum.
This makes a difference when squaring the delta function.
For a squared delta function of energy, one can use the
standard rule of evaluating $\delta(0)$ in terms
of the elapsed conjugate time variable, as is done in the usual textbook
``golden rule'' calculation.  For a delta function of absolute value of
momentum, it is not so clear what to use as the corresponding
conjugate variable when taking the square, and this appears to be the
root of the difficulties in the earlier calculations of refs [1], [2],
and [4].

In Sec. 3 we compare the Hornberger--Sipe and Di\'osi results
and show that they are the same; hence
Di\'osi's 1995 calculation appears to be the first giving the correct
result for the collisional decoherence rate.
In Sec. 4 we give an independent check of the corrected expression for
collisional decoherence, by using the corresponding master equation to
evaluate the scattering-induced translational Brownian diffusion of a hard
sphere in the geometric scattering limit. Planck's constant cancels out
in this calculation, and so
the result obtained this way can be directly compared with the
classical Brownian diffusion of a hard sphere, and the two calculations are
in precise agreement.

\bigskip
\centerline{\bf 2.~~ Calculation using a squared energy delta function}
\bigskip
To keep this section concise, we will use the notation of Hornberger and
Sipe, and give just a brief summary of their calculation up to the point
where our treatment begins to differ from theirs.  We consider a Brownian
particle in a bath of $N$ scattering particles of mass $m$,
contained in normalization
volume $\Omega$. In the dilute case,  the scatterings of the bath particles
from each other can be neglected, and their scatterings from the Brownian
particle are independent of one another.  Then the effect of the $N$
bath particles is obtained by considering the effect of
a single bath particle,
and multiplying by $N$ at the end of the calculation.  For a single
scattering, the effect of the collision is to change the density matrix
$\rho_0({\bf R}_1,{\bf R}_2)$ to
$$\rho({\bf R}_1,{\bf R}_2) =\eta({\bf R}_1,{\bf R}_2)
\rho_0({\bf R}_1,{\bf R}_2)~~~,\eqno(1)$$
with the factor $\eta({\bf R}_1,{\bf R}_2)$ given by
$$\eta({\bf R}_1,{\bf R}_2)=
{\rm tr}_{\rm bath}\{e^{-i{\bf p}\cdot {\bf R}_2/\hbar}{\cal S}_0^{\dagger}
e^{i{\bf p} \cdot ({\bf R}_2-{\bf R}_1)/\hbar} {\cal S}_0
e^{i{\bf p}\cdot {\bf R}_1/\hbar} \rho^{\rm bath} \}~~~. \eqno(2)$$
Here ${\cal S}_0$ is the scattering matrix, and $\rho^{\rm bath}$
corresponds to an ensemble momentum space weighting
$$\int d{\bf p} \mu({\bf p})~~~,\eqno(3a)$$
where for a thermal ensemble with $\beta = (kT)^{-1}$ one has
$$\mu({\bf p})=\left({\beta \over 2 \pi m}\right)^{3/2} e^{-\beta {\bf p}^2
/(2m)}~~~.\eqno(3b)$$
Substituting ${\cal S}_0=1+i{\cal T}_0$, evaluating the bath trace in a
momentum basis, inserting a complete set of intermediate states, changing
from box to continuum normalization, and
using the unitarity relation $i({\cal T}_0-{\cal T}_0^{\dagger})
=-{\cal T}_0^{\dagger} {\cal T}_0$, Hornberger and Sipe show that Eq.~(2)
takes the form
$$\eta({\bf R}_1,{\bf R}_2)=
\int d{\bf p} \mu({\bf p})
\left[1-{(2\pi \hbar)^3\over \Omega} \int d{\bf p}^{\prime}
(1-e^{i({\bf p}- {\bf p}^{\prime}) \cdot ({\bf R}_1-{\bf R}_2)/\hbar})
{\bf |}\langle {\bf p}^{\prime}|{\cal T}_0| {\bf p} \rangle {\bf |}^2
\right].~~~\eqno(4a)$$
This equation, which is Eq.~(51) of Hornberger and Sipe, and without
the weighting
over $\mu({\bf p})$ also corresponds to the first line of Eq.~(2.11) of
Gallis and Fleming, will be the starting point for our analysis.

Denoting the elapsed time in the scattering process by $T$, so that
Eqs.~(1) and (4a) implicitly refer to time $T$,
and recalling that $1=\int d{\bf p} \mu({\bf p})$,
we can rewrite Eqs.~(1) and (4a) as
$$\eqalign{
&\rho({\bf R}_1,{\bf R}_2;T)-\rho({\bf R}_1,{\bf R}_2;0)
= (\eta({\bf R}_1,{\bf R}_2;T)-1) \rho_0({\bf R}_1,{\bf R}_2)  \cr
=&- \rho_0({\bf R}_1,{\bf R}_2)\int d{\bf p} \mu({\bf p})
{(2\pi \hbar)^3\over \Omega} \int d{\bf p}^{\prime}
(1-e^{i({\bf p}-{\bf p}^{\prime}) \cdot ({\bf R}_1-{\bf R}_2)/\hbar})
{\bf |}\langle {\bf p}^{\prime}|{\cal T}_0| {\bf p} \rangle {\bf |}^2 ~~~.\cr
}\eqno(4b)$$
Our next task will be to evaluate the squared matrix element appearing
in the integrand of Eq.~(4b).

The general ${\cal T}_0$ matrix element
$\langle {\bf q}_2|{\cal T}_0|{\bf q}_1 \rangle$
can as usual be expressed in terms of the scattering
amplitude $f({\bf q}_2,{\bf q}_1)$ and an energy-conserving delta
function,
$$\eqalign{
\langle {\bf q}_2|{\cal T}_0|{\bf q}_1 \rangle=&
{1\over 2 \pi \hbar m}\delta(E_2-E_1) f({\bf q}_2,{\bf q}_1) \cr
=& {1\over 2 \pi \hbar q_2} \delta(q_2-q_1) f({\bf q}_2,{\bf q}_1)~~~.\cr
}\eqno(5)$$
Instead of using the second line of Eq.~(5) to form the square of the
${\cal T}_0$ matrix element,  we will use the first line, using the
second line only {\it after} the delta function of zero energy argument
has been evaluated.  Thus, we have
$${\bf |}\langle {\bf q}_2|{\cal T}_0|{\bf q}_1 \rangle {\bf |}^2
={1\over (2 \pi \hbar m)^2 } \delta^2(E_2-E_1)
{\bf |} f({\bf q}_2,{\bf q}_1) {\bf |}^2~~~.\eqno(6)$$
Using the Fourier representation for the energy delta function,
$$
\delta(E_2-E_1)={1\over 2 \pi \hbar}\int_{-\infty}^{\infty} dt
\exp[i (E_2-E_1) t/\hbar]~~~,\eqno(7a)$$
we find
$$\eqalign{
\delta^2(E_2-E_1) =&\delta(E_2-E_1) \delta(0)  \cr
=&\delta(E_2-E_1) {1\over 2 \pi \hbar} \int dt\cr
= &\delta(E_2-E_1) {T \over 2 \pi \hbar}\cr
=& {m\over q_2} {T \over 2 \pi \hbar} \delta(q_2-q_1)~~~,\cr
}\eqno(7b)$$
where in the final line we have
converted the energy delta function to a delta function of the absolute
value of the three-momentum.
Here $T$ is the elapsed time interval, which we assume to be longer 
than the time for a single scattering, but still short compared to 
the characteristic decoherence time of the Brownian particle in the 
$N$-particle bath.  
 
 Replacing ${\bf q}_2, {\bf q}_1$ by
${\bf p}^{\prime}, {\bf p}$ respectively, substituting Eqs.~(7b) and (5)
into Eq.~(4b), and writing $ d{\bf p}^{\prime} = d{\hat n}
(p^{\prime})^2 dp^{\prime}$, with $d{\hat n}$ a solid angle
differential, we can immediately integrate out the delta function
of the absolute value of momentum. Using
$${(2\pi \hbar)^3\over \Omega} {1\over (2 \pi \hbar m)^2} p^2 {m\over p}
 {T \over 2 \pi \hbar} =  {T \over \Omega} {p \over m}~~~,\eqno(8a)$$
we thus get
$$\eqalign{
\rho({\bf R}_1,{\bf R}_2;T)-\rho({\bf R}_1,{\bf R}_2;0)
=&-\rho_0({\bf R}_1,{\bf R}_2)  \cr
&\times {T \over \Omega}  \int d{\bf p} \mu({\bf p})
{p \over m}     \int d{\hat n}
(1-e^{i({\bf p}- p{\hat n}) \cdot ({\bf R}_1-{\bf R}_2)/\hbar})
{\bf |}f( p {\hat n},{\bf p}){\bf |}^2 ~~~. \cr
}\eqno(8b)$$

Multiplying by $N$ to take account of the fact that each of the $N$ bath
particles makes a contribution equal to Eq.~(8b),
denoting the bath density $N/\Omega$ by $n$,
dividing by $T$, and
finally passing to the limit of small $T$, we get the result
$${\partial \rho({\bf R}_1,{\bf R}_2;t) \over \partial t}
= -F({\bf R}_1-{\bf R}_2) \rho({\bf R}_1,{\bf R}_2;t) ~~~,\eqno(9a)$$
with
$$ F({\bf R}_1-{\bf R}_2) =
n  \int d{\bf p} \mu({\bf p})
{p \over m}     \int d{\hat n}
(1-e^{i({\bf p}- p{\hat n}) \cdot ({\bf R}_1-{\bf R}_2)/\hbar})
{\bf |}f( p {\hat n},{\bf p}){\bf |}^2 ~~~.\eqno(9b)$$
This is the form of the final result for the decoherence-induced master
equation given in Eq.~(55), and in the
unnumbered immediately preceding equation, of Hornberger and Sipe [4].
By defining $\nu(p)$ by
$$
\mu({\bf p} ) d{\bf p} = {\nu(p) dp d{\hat s} \over 4 \pi}~~~,\eqno(10a)$$
with $d{\hat s}$ a second solid angle differential, so that
$\int_0^{\infty} dp \nu(p)=1$, Hornberger and Sipe also rewrite Eq.~(9b)
in the equivalent form (after a relabeling of the integration variables)
$$ F({\bf R}) =
n  \int_0^{\infty} dq  \nu(q )
{q \over m}  \int {   d{\hat n}_1 d{\hat n}_2  \over 4 \pi}
(1-e^{iq({\hat n}_1- {\hat n}_2) \cdot{\bf R}/\hbar})
{\bf |}f( q {\hat n}_2, q {\hat n}_1){\bf |}^2 ~~~.\eqno(10b)$$

In the limit of large {\bf R}, the exponential term in Eqs.~(9b) and
(10b) averages to zero provided that  $ {\hat n}_2 \neq  {\hat n}_1$,
while in the forward direction  $ {\hat n}_2 =  {\hat n}_1$ the integrand
in Eqs.~(9b)  and (10b) vanishes for all {\bf R}.
Hence one has
$$\eqalign{
F({\bf R}\to {\bf \infty}) =&
n  \int_0^{\infty} dq  \nu(q )
{q \over m}  \int_{ {\hat n}_2 \neq  {\hat n}_1} {   d{\hat n}_1 d{\hat n}_2  \over 4 \pi}
{\bf |}f( q {\hat n}_2, q {\hat n}_1){\bf |}^2\cr
=& n  \int_0^{\infty} dq  \nu(q ) {q \over m} \sigma(q)~~~,\cr
}\eqno(10c)$$
with $\sigma(q)$ the total cross section (excluding a possible delta function
contribution to the forward diffraction
peak).  In other words, the large {\bf R} asymptote of $F({\bf R})$ is
the thermal ensemble average $\langle n v \sigma \rangle_{\rm AV}$
of the non-forward scattering rate $n (q/m) \sigma(q)$.  Correspondingly,
from Eqs.~(4b) and (8b), as modified by multiplication by the factor $N$, 
we see that $\eta({\bf R}_1,{\bf R}_2;T)$
$= 1-T F({\bf R}_1-{\bf R}_2)$ has the large {\bf R} asymptote
$\eta({\bf \infty};T)=1-T \langle n v \sigma \rangle_{\rm AV}$.
Thus $\eta({\bf \infty};T)$ vanishes for $T$ equal to the inverse of the
averaged non-forward scattering rate, a result reminiscent of, but
not identical to, the condition   $\eta({\bf \infty})=0$
imposed by Hornberger and Sipe on the
single collision decoherence function in their Eq.~(45), on which they base
their method for evaluating the square of a delta function of the absolute
value of momentum.  (We emphasize, however, that in the calculation 
leading to Eq.~(9b) we have not fixed $T$ by imposing such a condition.)

\bigskip
\centerline{\bf 3.  Comparison of the Hornberger-Sipe and Di\'osi Results}
\bigskip
The calculation of Di\'osi includes effects of recoil of the Brownian
particle; we show in this section that in the limit of an infinitely heavy
Brownian particle, the results of Hornberger--Sipe and of Di\'osi are in
agreement.  When recoil is neglected,   Di\'osi's result is his
Eq.~(19), which reads
$${d \rho \over dt} = n_0 \int dE d\Omega_i d\Omega_f k^2
{d \sigma(\theta, E) \over d\Omega_f} \rho^{\cal E}({\bf k}_i)
\left( V_{{\bf k}_f {\bf k}_i} \rho  V_{{\bf k}_f {\bf k}_i}^{\dagger}
-{1\over 2} \{  V_{{\bf k}_f {\bf k}_i}^{\dagger}  V_{{\bf k}_f {\bf k}_i}
,\rho \} \right)~~~.\eqno(11a)$$
Substituting into Eq.~(11a) Di\'osi's Eq.~(20) (with the Brownian particle recoil
term dropped),
$$ V_{{\bf k}_f {\bf k}_i}=\exp(-i {\bf k}_{fi} \cdot \bf{q})
=\exp(-i({\bf k}_f-{\bf k}_i) \cdot \bf{q})~~~,\eqno(11b)$$
taking the matrix element of Eq.~(11a) between {\bf q} eigenstates
$\langle {\bf R}_1|$ and $|{\bf R}_2 \rangle$, and writing
$$  \rho({\bf R}_1,{\bf R}_2;t) =
\langle {\bf R}_1 |\rho(t)|{\bf R}_2 \rangle~~~,\eqno(11c)$$
we get
$${d \rho({\bf R}_1,{\bf R}_2;t) \over dt} =-n_0
\int dE d\Omega_i d\Omega_f k^2
{d \sigma(\theta, E) \over d\Omega_f} \rho^{\cal E}({\bf k}_i)
\left(1-e^{i({\bf k}_i - {\bf k}_f) \cdot ({\bf R}_1-{\bf R}_2) }  \right)
 \rho({\bf R}_1,{\bf R}_2;t)
~~~.\eqno(12)$$
Taking account of the fact that $dE=dk^2/(2m) =(k/m)dk$, together with
$d \sigma(\theta, E)/ d\Omega_f={\bf |} f {\bf |}^2$, where
$f$ is the scattering amplitude, along with $ \rho^{\cal E}({\bf k}_i) =
\mu({\bf k}_i)$ and some obvious relabelings of variables, one sees
that Eq.~(12) is identical to Eqs.~(9a) and (9b) that follow from the
analysis of Hornberger and Sipe.

Di\'osi also gives the expansion of
Eq.~(12) to leading order in ${\bf R} = {\bf R}_1 - {\bf R}_2$.  Working
now in the other direction, from Eq.~(10b), the simplest way to find the
leading
order ${\bf R}$ dependence is to note that $F({\bf R})$ is a rotationally
invariant function of ${\bf R}$. Hence it suffices  to evaluate the
average over the direction of ${\bf R}$.  Expanding out the exponential
in Eq.~(10b), we have
$$1-e^{iq({\hat n}_1- {\hat n}_2) \cdot {\bf R}/\hbar }
=1-iq({\hat n}_1- {\hat n}_2) \cdot{\bf R}/\hbar
+{1\over2} (q^2/\hbar^2) [({\hat n}_1-{\hat n}_2) \cdot {\bf R}]^2+...~~~.
\eqno(13a)$$
The average of $[({\hat n}_1-{\hat n}_2) \cdot {\bf R}]^2$
over the direction of {\bf R} is
$$\eqalign{
{1\over 3} R^2  \sum_i [({\hat n}_1-{\hat n}_2) \cdot \hat i]^2
=& {1\over 3} R^2 ({\hat n}_1-{\hat n}_2)^2 \cr
=&{2\over 3} R^2 (1-{\hat n}_1 \cdot {\hat n}_2)
= {4 \over 3} R^2 \sin^2(\theta/2)~~~,\cr
}\eqno(13b)$$
with $\theta$ the angle (the scattering angle) between ${\hat n}_1$
and ${\hat n}_2$.
Substituting Eqs.~(13a,b) into Eq.~(10b), we get
$$F({\bf R})=R^2 \Lambda~~~, \eqno (14a)$$
with
$$\eqalign{
\Lambda =& {2 \over 3} {n \over \hbar^2}
  \int_0^{\infty} dq  \nu(q )
{q \over m}q^2  \int {   d{\hat n}_1 d{\hat n}_2  \over 4 \pi}
\sin^2(\theta/2) {\bf |}f( q{\hat n}_2,q{\hat n}_1){\bf |}^2 \cr
=&{2 \over 3} {n \over \hbar^2}
\int d{\bf q} \mu({\bf q})  {q \over m}q^2 \int d{\hat n}_2
\sin^2(\theta/2) {\bf |}f( q{\hat n}_2,q{\hat n}_1){\bf |}^2 \cr
}~~~.\eqno(14b)$$
With the shifts in notation noted above, this equation for $\Lambda$
is identical to Eq.~(22) of Di\'osi, which gives what he terms the
diffusion parameter $D_{pp}$.
\bigskip
\centerline{\bf 4.~~Comparison of decoherence-based and classical}
\centerline{\bf ~~~~calculations of Brownian translational diffusion}
\bigskip
Let us now apply the result of Eqs.~(14a,b) to the quantum Brownian motion
of a Brownian particle of mass $M$ and radius $a$, with $a$ large enough
so that $pa>>1$ for
important bath particle momenta.  In this case the quantum scattering
differential cross section consists of two parts [9]:  an isotropic part,
with an integrated
cross section contribution of $\pi a^2$, and a forward diffraction peak,
again contributing cross section $\pi a^2$. Since the integrand of Eq.~(14b)
vanishes for forward scattering, the forward diffraction peak makes no
contribution. Thus we can evaluate the integrals by taking the scattering
amplitude to be a constant, $f( q {\hat n}_2, q {\hat n}_1) =F$,
with
$$\int d{\hat s} {\bf |} F {\bf |}^2= 4 \pi {\bf |} F {\bf |}^2= \pi a^2
~~~,\eqno(15a)$$
that is, with
$$ {\bf |} F {\bf |}^2 =a^2/4~~~.\eqno(15b)$$
Substituting into Eq.~(14b),
we then find that
$$\Lambda= {n \pi a^2 \langle q^2 v \rangle_{\rm AV} \over 3 \hbar^2}
~~~.\eqno(16)$$
Here we have defined
$$ \langle q^2 v \rangle_{\rm AV}
=\int_0^{\infty} dq \nu(q) q^2 {q \over m}
=4 (m/\pi)^{1\over 2} (2kT)^{3\over 2}~~~,\eqno(17a)$$
where we used the expression
$$\nu(q)=4 \pi q^2 \left( {\beta \over 2\pi m}\right)^{3\over 2}
e^{-\beta q^2 /(2m)}~~~, \eqno(17b)$$
which follows from the definitions of Eqs.~(3b) and (10a), to evaluate
the thermal average.

Substituting Eq.~(14a) into Eq.~(9a), we then get
$${\partial \rho({\bf R}_1,{\bf R}_2;t) \over \partial t}
= -\Lambda ({\bf R}_1-{\bf R}_2)^2 \rho({\bf R}_1,{\bf R}_2;t) ~~~,
\eqno(18a)$$
which with Eq.~(11c)  is equivalent to the operator equation
$${\partial \rho(t) \over \partial t} =
-\Lambda \sum_{j=1}^3 [R_j,[R_j,\rho(t) ]]~~~,\eqno(18b)$$
with $R_j$ denoting the Cartesian components of {\bf R}.
Adding the kinetic energy term to the differential equation for $\rho(t)$,
we get the total evolution equation
$${\partial \rho(t) \over \partial t} =
-{i\over \hbar} [H_{\rm kin}, \rho(t)]
-\Lambda \sum_{j=1}^3 [R_j,[R_j,\rho(t) ]]~~~,\eqno(19a)$$
with the kinetic Hamiltonian for the Brownian particle given by
$$H_{\rm kin}= \sum_{j=1}^3 {P_j^2 \over 2M}~~~. \eqno(19b)$$
Here $P_j$ is the momentum operator corresponding to the coordinate
operator $R_j$, so that $[R_j, P_k] = i \hbar \delta_{jk}$.

{}From Eqs.~(19a) and (19b), one  can calculate the mean squared coordinate
diffusion as a function of time, for a Brownian particle that starts
at ${\bf R}={\bf 0}$ at $t=0$ with zero drift velocity. This 
calculation in the one-dimensional
case is given in Adler [10],
by constructing a generating function for the trace of $\rho(t)$
multiplied by an arbitrary polynomial constructed from $R_j$ and $P_k$.
For the mean square coordinate deviation, the result on converting
to the present notation is (with no sum
implied over $j$)
$$\langle R_j^2 \rangle = {\rm tr} \rho(t) R_j^2
= {2 \Lambda \hbar^2 t^3 \over 3 M^2}~~~.\eqno(20a)$$
Substituting Eq.~(16) for $\Lambda$ and Eq.~(17a) for the thermal
average of $q^2 v$, we get finally (again with $j$ unsummed)
$$ \langle R_j^2 \rangle
=C (kT)^{3\over 2}n m^{1\over 2} a^2t^3/M^2~~~, \eqno(20b)$$
with
$$C={16\over 9} (2 \pi)^{1\over 2}~~~.\eqno(20c)$$
Note that the Planck constant $\hbar$ has dropped out of this result.
Hence the formula of Eqs.~(20b,c) is a classical result, and should be
recoverable by a purely classical calculation.

The formulas needed for a classical Brownian motion
evaluation of $\langle R_j^2 \rangle$
are summarized in a recent paper by Collett and Pearle [11].  Their
Eq.~(2.2) gives (again with $j$ unsummed)
$$\langle R_j^2 \rangle= {2 kT \xi t^3 \over 3 M^2}~~~,\eqno(21a)$$
with $\xi$ a viscosity factor, which for a sphere of radius $a$ in a
dilute bath is given by their Eq.~(2.5),
$$\xi={8\over 3} na^2 (2 \pi m kT)^{1\over 2}~~~.\eqno(21b)$$
Substituting Eq.~(21b) into Eq.~(21a) then gives a result identical
to Eqs.~(20b,c) above.  Since the result of Eqs~(20b,c) is directly
proportional to the normalization constant in the collisional decoherence
rate, this agreement gives added confirmation of the correctness of
Eq~(9b).

\centerline{\bf Acknowledgments and Addenda}
\bigskip
This work was supported in part by the Department of Energy under
Grant \#DE--FG02--90ER40542, and was done while the author was at the
Aspen Center for Physics.  I wish to Philip Pearle for alerting me
to reference [9], which was the initial stimulus for this investigation,
and Jonathan Halliwell for bringing
references [2] -- [4] to my attention. A related discussion of the
derivation of the decoherence master equation is given in a recent
article of Halliwell [12].  I also wish to thank Klaus Hornberger for
an informative email correspondence, and for noting a recent paper [13],
in which he extends the method of ref [4] to the case of a finite mass
Brownian particle.  Finally, Alexander Pechen has pointed out a recent
paper [14], giving further references, pertaining to the microscopic
derivation of the master equation for a quantum system interacting with
a dilute environment.
\vfill\eject
\centerline{\bf References}
\bigskip
\noindent
[1]  Joos E and Zeh H D (1985) {\it Z. Phys. B: Condens. Matt.}
{\bf 59} 223. \hfill\break
\bigskip
\noindent
[2] Gallis M R and Fleming G N (1990) {\it Phys. Rev. A} {\bf 42} 38.
\hfill\break
\bigskip
\noindent
[3] Dodd P J and Halliwell J J (2003) {\it Phys. Rev. D} {\bf 67} 105018.
See also the related recent article of Halliwell [10].
\hfill\break
\bigskip
\noindent
[4] Hornberger K and Sipe J E (2003) {\it Phys. Rev. A} {\bf 68} 012105.
See also the related recent article of Hornberger [11].
\hfill\break
\bigskip
\noindent
[5] Di\'osi L (1995) {\it Europhys. Lett.} {\bf 30} 63 (1995). \hfill\break\bigskip
\noindent
[6] Altenm\"uller T P, M\"uller R and Schenzle A (1997) {\it Phys. Rev. 
A} {\bf 56} 2959.\hfill\break
\bigskip
\noindent
[7] Vacchini B (2000) {\it Phys. Rev. Lett.} {\bf 84} 1374.\hfill\break
\bigskip
\noindent
[8] Hornberger K, Uttenthaler S, Brezger B, Hackerm\"uller L,
Arndt M and Zeilinger A (2003) {\it Phys. Rev. Lett.} {\bf 90} 160401.
\bigskip
\noindent
[9] Schiff L I (1968) {\it  Quantum Mechanics, 3rd ed}  (New York:
McGraw-Hill) pp 124-126.
\hfill\break
\bigskip
\noindent
[10] Adler S L (2005)  {\it J. Phys. A: Math. Gen.} {\bf 38} 2729.
\hfill\break
\bigskip
\noindent
[11] Collett B and Pearle P (2003) {\it Found. Phys.} {\bf 33} 1495.
\hfill\break
\bigskip
\noindent
[12] Halliwell J J (2006) ``Two Derivations of the Master Equation of
Quantum Brownian Motion'' to appear in {\it J. Phys. A: Math. Gen.}.
\hfill\break
\bigskip
\noindent
[13] Hornberger K (2006) ``Master equation for a quantum particle in
a gas'' arXiv:quant-ph/0607085.  \hfill\break
\bigskip
\noindent
[14] Pechen A (2006) ``White noise approach to the low density limit
of a quantum particle in a gas'' arXiv:quant-ph/0607134.\hfill\break
\bigskip
\noindent
\bigskip
\noindent
\bigskip
\noindent
\bigskip
\noindent
\bigskip
\noindent
\bigskip
\noindent
\bigskip
\noindent
\bigskip
\noindent
\bigskip
\noindent
\bigskip
\noindent
\bigskip
\noindent
\vfill
\eject
\bigskip
\bye